\documentstyle[preprint,revtex]{aps}
\def\bb{\begin{equation}}
\def\ee{\end{equation}}
\def\hf{\hbar_{\rm eff}}

\begin{document}

\draft
\begin{title}
Traces of a Quantum Anti Resonance in a Driven System
\end{title}
\author{\bf E. Eisenberg, R. Avigur and N. Shnerb}
\begin{instit}
Department of Physics, Bar-Ilan University, Ramat-Gan 52900, Israel
\end{instit}
\begin{abstract}
It has been previously shown, that classically chaotic kicked systems,
whose unperturbed spectrum posseses one energy scale, admits a quantum
anti-resonance (QAR) behavior. In this study we extend the
conditions under which which this QAR occurs for the case of a two-sided
kicked 1D infinite potential well. It is then shown by a perturbative
argument that this QAR effects the behavior of the equivalent driven well,
i.e., the number of periods needed to leave the initial state
has a sharp peak around the QAR. We give a numerical evidence that the
anti resonance persists even for large values of the perturbation
parameter.
\end{abstract}

\narrowtext
\newpage

The study  of ``quantum chaos",  i.e., understanding the  fingerprints of
classical  chaos  in  quantum  mechanics  \cite{book},  has  led  to  the
discovery of a variety of  new quantum-dynamical phenomena.  Several such
phenomena  occur  in  time-periodic  systems  described  by  the  general
Hamiltonian
\begin{equation}\label{H}
H = H_0 + H_1 f(t)\ ,
\end{equation}
where  $H_0$ is  some  time-independent Hamiltonian,  $H_1$ represents  a
perturbation, and $f(t)$ is periodic  with period $T$, $f(t+T)=f(t)$.  In
many  cases,  $f(t)$ is  chosen,  for  simplicity,  as a  periodic  delta
function,   $f(t)=\Delta_T(t)\equiv    \sum_{s=-\infty}^{\infty}   \delta
(t-sT)$,   giving   the   well-studied  class   of   ``kicked"   systems.
Representative  models   in  this  class   are  the  kicked   rotor  (KR)
\cite{rot,izr,hog,fish,kr3},  the kicked  Harper model  (KHM)
\cite{leb,lim,da1,bo1},   and    the   kicked
harmonic oscillator (KHO) \cite{da1,ber}.
In recent years a different kind of  phenomenon for time-periodic
systems has been studied: exactly {\it periodic} recurrences.
This phenomenon is defined, in general, by
\begin{equation}\label{QAR}
U^p = e^{-i\beta}\ ,
\end{equation}
where   $U$  is   the  one-period   evolution  operator   for  (\ref{H}),
$e^{-i\beta}$ is  some constant phase factor  (a $c$ number), and  $p$ is
the smallest positive  integer for which (\ref{QAR})  is satisfied.  Thus
$pT$ is the recurrence period.

The class  of systems considered  are the two-sided kicked  rotors (TKRs)
\cite{us,des}, defined by the Hamiltonian
\begin{equation}\label{2kr}
H = \frac{L^2}{2I} +
    \hat k V(\theta )\sum _{s=-\infty}^{\infty}
    (-1)^s \delta \left ( t-\frac{sT}{2} \right ) \ ,
\end{equation}
where $I$  is the moment of  inertia, $\hat k$ is  the kicking parameter,
$T$  is the  time period,  and  $V(\theta )$  is a  general periodic  and
analytic function of the angle $\theta$.  Two-sided kicking perturbations
such  as in  (\ref{2kr})  were considered  in  several physical  contexts
\cite{gy}   as   approximations    of   sinusoidal   driving   potentials
corresponding to  ac electromagnetic fields.   By increasing $\hat  k$ in
the classical  TKR, one observes  the typical transition from  bounded to
global  chaos \cite{us},  as in  the KR  case.  The  quantum dynamics  is
governed, as  usual, by the evolution  operator $U$ in one  period, e.g.,
from $t=-0$ to $t=T-0$:
\begin{equation}\label{U}
U=e^{-i\tau \hat n ^2}e^{ikV(\theta )}
  e^{-i\tau \hat n ^2}e^{-ikV(\theta )} \ ,
\end{equation}
where $\hat n  \equiv L/\hbar =-id/d\theta$, $\tau  \equiv \hbar T/(4I)$,
and  $k\equiv \hat  k /\hbar$.

A most distinctive feature of the quantum TKR is that $U$ becomes the
identity operator  for $\tau=2\pi m$, an integer  (since the  operator $\exp
(-i\tau \hat n ^2)$ in (\ref{U}) is clearly the identity in this case).
This implies  {\it exactly}  periodic recurrences (with  period 1)  of an
arbitrary wave-packet  \cite{us}. This  phenomenon is refered to as
quantum antiresonance (QAR).

The previously described QAR was based on the fact that the
unperturbed evolution between succesive kicks was described by the
identity operator such that the opposite sign kicks are cancelled.
It is therefore clear that the same phenomenon will occur
for a two sided kicked 1D infinite potential well.
In the present study, we show the existence of a new anti-resonance phenomena
for the case of the linearly kicked one dimensional potential well.
The new QAR occurs when the period of the kicks is {\it half} of the
period needed for the more general QAR, as described in Eq. (\ref{U}).
Thus, this QAR is, in some sence, a ``period halfing'' of the former, and
specific to this special case, as we will discuss below.
It is then shown that this new QAR, namely the one which ocuurs
when a complete period of the perturbation is compatable
with the level spacing frequencies, affects the behavior
of the corresponding driven system. In particular, in the region of
the QAR condition, the driven system becomes nearly periodic.
This may be considered as a 'trace' of the exact QAR in the kicked
system. Numerical edivence for this effect are presented.

It is interesting to note that such a model system can in fact be
realized experimentally.
Modern semiconductor technology has enabled the fabrication of 1D
quantum wells \cite{hol1}. Such a quantum well is fabricated by varying
the alloy composition in a compound semiconductor like $Al_x Ga_{1-x} As$
along one dimension. Conduction electrons in such structures experience an
arbitrarily shaped effective potential in the growth direction while
remaining free in the perpendicular plane. Quantum wells are
typically $200 - 300$ meV deep with level spacing $\Delta E$
between several meV and $150$ meV. These systems are of special
interest since they can be treated in means of pure quantum
mechanical considerations while they are still experimentally
accessible.
Recently, there has been intereset in the behavior of such systems
under the influence of an electromagnetic field
\cite{anals,rcb,holt,reich1}. The quantum well
structure can be considered as an analogue of an 1D atom, and thus
a study of the driven well may help us learn about the interaction of
atoms and high-field electromagnetic radiation. In the
region where the electric field is strong with respect to
the level spacing of the well, one obtains a system in
which to study non-perturbative effects in light-matter interaction.
The QAR behavior is therefore experimentally accessible.
One should expect a sharp anti-peak in the absorption spectrum
of the quantum well.

Consider the Hamiltonian
\bb
H = \frac{P^2}{2m} + \lambda \bar X  {\sum_{m=-\infty}^\infty}
\{\delta(t/T-m) - \delta(t/T-m+1/2)\},
\ee
defined in the infinite well $x\in [0,L]$.
We use the dimensionless form, defined by the transformation
$\tau=2\pi t/T$, and $X=\bar X/L$ and obtain the Schrodinger
equation
\bb
i\frac{d\psi}{d\tau}={\cal H}\psi =\left\{ -\frac{\hf}{2} \partial_x^2
+ \frac {\beta}{\hf} X
{\sum_{m=-\infty}^\infty}
\{\delta(\tau-2\pi m) - \delta(\tau-2\pi(m+1/2))\}\right\}\psi,
\ee
where $\hf=\hbar/(m\omega L^2)$ and $\beta=\lambda/(m\omega^2L)$.
The evolution  operator for  takes the form
\bb
U = \exp(iF) = e^{i\pi\frac{\hf}{2}\partial_x^2}
e^{-i\frac{\beta }{\hf} X }
e^{i\pi\frac{\hf}{2} \partial_x^2} e^{i\frac{\beta }{\hf} X }.
\ee

Evidently, since the eigenvalues of the operator $-\partial_x^2$
are of the form $n^2 \pi^2$ with $n$ integer,
if $\hf$ takes the values $\hf = 4k/\pi^2$ ($k$ integer),
the free evolution of the system between two kicks turns out to be
the identity operator, and the fact that the two kicks are with oposite sign
implies that the whole evolution  operator is also the identity. This
is the usual quantum  anti-resonance.
However, we now show that for this special case, a QAR exist
also when  $\hf = 2k/\pi^2$. In order to do that, we point out that
\bb
e^{i\pi \frac{\hf}{2} \partial_x^2} e^{-i\frac {\beta }{\hf} X} =
 e^{-i\frac {\beta }{\hf} X_I}  e^{i\pi\frac{\hf}{2} \partial_x^2},
\ee
where $(X_I)_{mn}  =  X_{mn}  e^{-i\pi \frac{\hf}{2}\pi^2 (m^2-n^2)}$.
since $ X_{mn}$ takes the form
\bb
X_{mn}=\left\{
\begin{array}{lr} -\frac{8mn}{\pi^2(m^2-n^2)^2} & {\rm m+n\ odd}\\
			0			&{\rm m+n\ even,m \neq n} \\
                       1/2                      &{m=n}
\end{array}\right.
\ee
it is clear that $(X_I)_{mn} = -X_{mn} + {\bf 1}$, where $ {\bf 1}$ is the
unit metrix, the whole evolution operator $U$ is then
\bb
U=e^{-iF}=e^{-i\beta/\hf},
\ee
which is the identity operator up to a constant  phase.
Thus, one obtains a ``period halfing'' of the usual
quantum anti-resonance.

We now turn to study the driven version of this problem.
This system is classically chaotic \cite{reich1} as shown in Figure 1.
Its quantum behavior is described
by the dimensionless Schrodinger equation
\bb
i\frac{d\psi}{d\tau}={\cal H}\psi =\left\{ -\frac{\hf}{2} \partial_x^2
+ \frac {\beta}{\hf} X\cos(\tau)\right\}\psi.
\ee
Clearly, due to the continuous character of the perturbation, the exact QAR
described above is not possible for this Hamiltonian. However, as we are
now to show, traces of the exact periodicity found in the kicked system
in case of QAR are seen in the driven system as well.
In fact, kicked systems are widely used as an approximation to
cosinusoidally driven perturbations. This can be understood in terms of the
relation
\bb
\sum_{n=-\infty}^{\infty}\biggl[\delta
\biggl({t\over T}-n\biggr) - \delta\biggl(
({t\over T}-(n+{1\over 2})\biggr)\biggr]\,
=\,4\sum_{n=1}^{\infty}\cos[(2n-1)\Omega t]\,,
\ee
through which, it is clear that in the limit where one may neglect the
effect of the higher frequencies, the kicks are essentially the same
as the $\cos$.

It turns out that under the condition $\hf=2k/\pi^2$ the QAR manifests
itself in the perturbative expansion of the time-dependent evolution
operator. This can be shown as follows.
The transition probability (to each order in $\beta$) for a complete period
is given in terms of a time-ordered integral of the form
\bb
\label{pert}
A^{(k)}(m\to n)\sim(\beta/\hf)^k T\int_0^{2\pi}dt_1\ldots\int_0^{2\pi}dt_k
X_I(t_1)\ldots X_I(t_k)\cos(t_1)\ldots\cos(t_k) .
\ee
The general form of $X_I(t)$ is
\bb
(X_I)_{mn}  =  X_{mn}  e^{-i\tau \frac{\hf}{2}\pi^2(m^2-n^2)},
\ee
and thus, (\ref{pert}) takes the form
\begin{eqnarray}
A^{(k)}(m\to n)&\sim&(\beta/\hf)^k \sum_{j_1,\ldots j_{k-1}}
X_{mj_1}X_{j_1j_2}\ldots X_{j_{k-1}n} \nonumber \\
& &T\int_0^{2\pi}dt_1\ldots\int_0^{2\pi}dt_k
\exp(-it_1\frac{\hf}{2}\pi^2(m^2-j_1^2)) \\
& &\exp(-it_2\frac{\hf}{2}\pi^2(j_1^2-j_2^2))\ldots
\exp(-it_k\frac{\hf}{2}\pi^2(j_{k-1}^2-n^2))
\cos(t_1)\ldots\cos(t_k) .\nonumber
\end{eqnarray}
In the anti resonance case, $\hf=2k/\pi^2$, all the frequencies in the
integrals are integers,and thus since the integral is over a complete period
the only contribution comes from the zero mode terms.
It is easy to see that even in the most 'soft' case, i.e. $k=1$,
the first contribution to the transition amplitudes is at least of order
$\beta^3$ (if the ground states is populated).
This can be seen due to the fact that to order
$\beta^k$ the integrand related to the transition
$m\to n$ contains the frequencies
$\omega_{mn}\pm 1\pm1\ldots \pm1\quad[{\rm k\ times}]$.
Since the smallest frequency is $\omega_{12}=3\alpha$, the first
and second order terms do not have any zero frequency components.
Thus the integral over a whole period vanishes.

In order to confirm the above predictions, we have solved the
time-dependent Schr\"odinger equation numerically, using a quality
control Runge-Kutta method, for various values of $\beta$ and
$\hf$ in the anti-resonance region. The results are
shown in Figure 2. We plot the inverse of the probability to leave the
initial state after one period. This quantity describes
approximately the amount of time needed to leave the initial state.
It is interesting to note that the QAR traces
persist even for large values of $\beta$, up to $\beta=2$. However,
as $\beta$ increases, the position of the anti resonance is slightly
shifted.

In conclusion, we have extended the concept of QAR for a case in which
the entire period of the perturbation is in resonance with the
level spacing frequencies. It has been shown by theoretical arguments and
numerical results that this effect persist for driven systems as well.
The conditions under which this effect occur can be realized in
experiments on quantum wells in far infra-red radiation.

\figure{Stroboscopic map of phase space trajectories for the linearly driven
	well, for $\beta=0.01$, avd several initial conditions.}
\figure{The inverse of $\Delta$, the probability to escape from
	the initial state after one period of the driving force,
	as a fuction of $\hf$, for various values of $\beta$.
	(a)$\beta$=0.1; (b)$\beta$=1.0; (c)$\beta$=2.0.}

\begin{references}

\bibitem{book} See, for example, M.   Gutzwiller, {\it Chaos in Classical
and Quantum Mechanics} (Springer,  Berlin, 1990), and references therein.
\bibitem{rot}  G.  Casati,  B.  V.   Chirikov, F.   M.  Izrailev,  and J.
Ford, in  {\it Stochastic Behavior  in Classical and  Quantum Hamiltonian
Systems}, Lecture Notes in Physics, Vol.  93, edited by G.  Casati and J.
Ford (Springer, Berlin, 1979); B.  V.  Chirikov, F.  M.  Izrailev, and D.
L.   Shepelyansky,  Sov.   Sci.   Rev.  {\bf  2C},  209  (1981);  F.   M.
Izrailev, Phys.  Rep. {\bf 196}, 299 (1990).
\bibitem{izr} F.  M.   Izrailev and D.  L.   Shepelyanskii, Theor.  Math.
Phys. {\bf 43}, 553 (1980).
\bibitem{hog} T.  Hogg and B.  A.  Huberman, Phys.  Rev.  Lett. {\bf 48},
711 (1982); Phys.  Rev.  A {\bf 28}, 22 (1983).
\bibitem{fish} S.   Fishman, D.  R.   Grempel, and R.  E.   Prange, Phys.
Rev.  Lett. {\bf 49}, 509 (1982); Phys.  Rev.  A {\bf 29}, 1639 (1984).
\bibitem{kr3} G.  Casati,  I.  Guarneri, and D.   L.  Shepelyansky, Phys.
Rev.  Lett. {\bf 62}, 345 (1989).
\bibitem{leb} P.  Leboeuf, J.  Kurchan, M.  Feingold, and D.  P.  Arovas,
Phys.  Rev.  Lett. {\bf 65}, 3076 (1990); {\it ibid.}, Chaos {\bf 2}, 125
(1992); F.  Faure and P.  Leboeuf,  in Proceedings of the Workshop ``From
Classical to  Quantum Chaos", G.   F.  Dell'Antonio, S.  Fantoni,  and V.
R.  Manfredi, editors, Trieste, 21-24 July 1992.
\bibitem{lim} R.  Lima and D.  Shepelyansky, Phys.  Rev.  Lett. {\bf 67},
1377 (1991).
\bibitem{da1} I. Dana, Phys. Rev. Lett. {\bf 73}, 1609 (1994).
\bibitem{bo1} F.   Borgonovi and D.  Shepelyansky,  Europhys.  Lett. {\bf
29}, 117 (1995).
\bibitem{ber} G.   P.  Berman,  V.  Yu.  Rubaev,  and G.   M.  Zaslavsky,
Nonlinearity {\bf 4}, 543 (1991).
\bibitem{us} E.  Eisenberg and N.  Shnerb,  Phys.  Rev.  E {\bf 49}, R941
(1994).
\bibitem{des} I.  Dana, E.  Eisenberg, and N.  Shnerb, Phys.  Rev.  Lett.
{\bf 74}, 686 (1995) and submitted.
\bibitem{gy}  Y.  Gu  and J.   M.  Yuan,  Phys.  Rev.   A {\bf  36}, 3788
(1987); G.  Casati, I.  Guarneri, and G.  Mantica (to be published).
\bibitem{hol1} See, for example, C. Weisbuch and B. Vinter, {\it Quantum
Semiconductor Structures}, Academic Press, San Diego 1991.
\bibitem{anals} H.P. Breuer and M. Holthaus, Ann. Phys. (NY) {\bf 221}, 249
	(1991).
\bibitem{rcb} B. Birnir, B. Galdrikian, R. Grauer and M. Sherwin, Phys. Rev.
	{\bf B47}, 6795 (1993).
\bibitem{holt} M. Holthaus, in K. Ikeda (ed.), {\it Quantum and Chaos,
	How Incompatible?}, Prog. Theor. Phys. Supp. {\bf 116} 1994, p. 417.
\bibitem{reich1} W.A. Lin and L.E. Reichel, Physica {\bf D19}, 145 (1986).
\end{references}
\end{document}